\documentclass[11pt,english,showkeys,showpacs,titlepage]{revtex4-2}
\usepackage[T1]{fontenc}
\usepackage[latin9]{inputenc}
\usepackage{float}
\usepackage{textcomp}
\usepackage{mathtools}
\usepackage{multirow}
\usepackage{amsmath}
\usepackage{amssymb}
\usepackage{graphicx}
\usepackage{wasysym}
\usepackage{esint}
\usepackage[all]{xy}
\usepackage[unicode=true,pdfusetitle,
 bookmarks=true,bookmarksnumbered=false,bookmarksopen=false,
 breaklinks=false,pdfborder={0 0 1},backref=false,colorlinks=false]
 {hyperref}
\usepackage{xcolor}
\usepackage[english]{babel}
\makeatletter

\usepackage{babel}

\newcommand{\xyR}[1]{%
\makeatletter
\xydef@\xymatrixrowsep@{#1}
\makeatother
} 

\newcommand{\xyC}[1]{%
\makeatletter
\xydef@\xymatrixcolsep@{#1}
\makeatother
} 
\makeatother

\begin{document}
\title{Triple  Charmonium Production in pQCD.}
\author{B. Blok}
\email{blok@physics.technion.ac.il}

\author{J. Mehl}
\email{yonatanm@campus.technion.ac.il}

\address{Department of Physics, Technion -- Israel Institute of Technology,
Haifa, Israel}
\begin{abstract}
We study the role of $1\rightarrow2$ and  $1\rightarrow3$ processes in triple charmonium production in the leading logarithmic approximation (LLA).
We see that 
the ratio of effective cross sections of TPS and DPS  only moderately depends on charmonium transverse momenta,
but the total DPS and TPS cross sections each separately may have rather strong dependence on charmonia transverse momenta in the 
central kinematics region that can be studied experimentally.
\end{abstract}
\maketitle
\section{Introduction.}
\par 
The multiparton interactions attracted a lot of attention in recent years both experimentally and theoretically.
In particular, the theory of double parton scattering (DPS) in QCD was the subject of 
 intensive developments in recent years. The first work on DPS were
done in the early 80s \citep{treleani,mekhfi}, and the first detailed experimental 
observations of DPS were done in Tevatron. Recently new detailed experimental
studies of DPS were carried  at LHC  and the new theoretical formalism based
on pQCD was developed \citep{GS1,Blok2011,diehl1,GS2,Blok2012,Diehl2016,Blok2014,diehl3,Manohar2012}.
In these works the fundamental role of parton correlations in DPS
scattering was realized and estimated and new physical objects to
study these correlations - two particle Generalized Parton Distributions
($_{2}GPD$s) were introduced.  These developments lead to much better understanding  of the experimental data on DPS 
production, like two dijet production, weak  bozon +  dijet, same sign WW etc . 
\par There  are still significant problems in comparing theory and experimental data, especially  in charmonium production in central kinematics,
where the measured DPS cross section is 2-4 times larger than the theoretically predicted one.
\par Note  however that  the contradictions between the theory based on pQCD and experimental data arise if one uses standard cross sections for single charmonium production, and model independent parameters for the calculation of effective cross section in the DPS mean field theory that can be extracted from HERA measurements \cite{Blok:2017alw,Blok2011,Blok2014} .  
\par The theory of both single and double charmonium production has considerable uncertainties \cite{Stasto}. Consequently, the double charmonium production is conventionally  described 
by a mean field formula, with the effective parameter  adjusted  in such a way that it gives experimentally correct DPS cross sections.
Such an approach was taken in recent  work  on  DPS charmonium production \cite{d1,dEnterria:2017yhd,Shao:2019qob}. 
\par The $1\rightarrow 2$ processes were shown to give  significant contributions to inclusive cross sections, leading to
effective cross sections depending on high transverse momenta. The ratio of mean field contribution  and of  the contribution  due to $1\rightarrow 2$ processes is usually  parametrised  by the parameter $R_{DPS}$
(defined exactly below). This $R_{DPS}$  is a product of a geometric factor and perturbative QCD term, both 
are independent of the value of the mean field approximation parameter.
\par Recently a new significant development in the study of  multiparton interactions occurred,
the observation by CMS of the  triple charmonium production. This production could only be explained 
by significant three particle scattering  (TPS) contributing to this process \cite{CMS:2023}, see also \cite{gaunt3} for comments on this measurement.
However, this process was studied theoretically only in mean field approximation
\cite{d1,dEnterria:2017yhd,Shao:2019qob}.
\par Recall that the DPS and TPS are conventionally parametrised  by  $\sigma_{eff;DPS}$, $\sigma_{eff;TPS}$ :
 \begin{equation}
\sigma_{TPS}=\frac{\sigma_{1}\sigma_{2}\sigma_{3}}{\sigma_{eff;TPS}^{2}}\label{1}
\end{equation}
and
\begin{equation}
\sigma_{DPS}=\frac{\sigma_{1}\sigma_{2}}{\sigma_{eff;DPS}}
\end{equation}
Here  $\sigma_i$, $i=1,2,3$ are  the cross sections of hard processes.
In the mean field approximation \cite{d1,dEnterria:2017yhd}, that was used in the pQCD analysis of TPS charmonium production  \cite{Shao:2019qob}:
\begin{equation}
\sigma_{eff;TPS}/\sigma_{eff:DPS}\sim 0.85\label{ratio}.
\end{equation}
This ratio is independent of the charmonium transverse momenta. The parameters of mean field approximation are adjusted
as to reproduce experimental  data on DPS charmonium production, but the ratio (\ref{ratio})  is independent of this adjusted parameter.
\par  In this paper we shall study the effect of $1\rightarrow 2$ and $1\rightarrow 3$  processes on the TPS cross section in the LLA approximation.
We shall see that the inclusion of $1 \rightarrow 2/3$ processes leads to a dependence on the transverse momenta 
$p_T$ 
for the  ratio (\ref{ratio}). It will be interesting to know how this dependence will influence the experimental data 
on TPS production and separation of TPS and DPS  produced charmonia.

 \par The paper is organized in the following way: In section \ref{sec:1to2} we remind the reader how to calculate $1\rightarrow 2$ processes and show the results for  $1\rightarrow 3$  process calculation. In section \ref{sec:Different} we give explicit formulae for different contributions to DPS and TPS, and in section \ref{sec:calculations} obtain the 
 corresponding analytical expressions, necessary for numerical calculations. In section \ref{sec:numeric} 
 we present  the numerical estimates of the cross sections and  give  the conclusions. In Appendix 
 \ref{sec:Computation-of-the} we give some details of the calculation of $1\rightarrow 3 $ processes in pQCD,

\section{$1\rightarrow 2 $ and $1\rightarrow 3 $ processes\label{sec:1to2}}
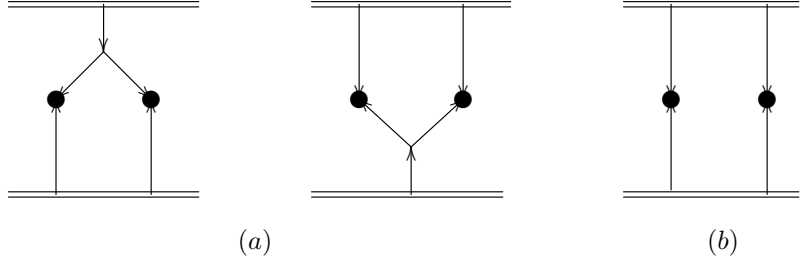
\begin{figure}
\[
\xymatrix{*=0{}\ar@{=}[rrrr] &  & *=0{}\ar@{->}[d] & *=0{} & *=0{} &  & *=0{}\ar@{=}[rrrr] & *=0{}\ar@{->}[dd] &  & *=0{}\ar@{->}[dd] & *=0{} &  & *=0{}\ar@{=}[rrrr] & *=0{}\ar@{->}[dd] &  & *=0{}\ar@{->}[dd] & *=0{}\\
\xyR{1pc}\xyC{1pc} &  & *=0{}\ar@{->}[dl]\ar@{->}[dr]\\
 & *=0{\newmoon} &  & *=0{\newmoon} &  & {\ \ } &  & *=0{\newmoon} & {\ } & *=0{\newmoon} &  & {\ \ } &  & *=0{\newmoon} & {\ } & *=0{\newmoon}\\
 &  &  &  &  &  &  &  & *=0{}\ar@{->}[ul]\ar@{->}[ur]\\
*=0{}\ar@{=}[rrrr] & *=0{}\ar@{->}[uu] &  & *=0{}\ar@{->}[uu] & *=0{} &  & *=0{}\ar@{=}[rrrr] &  & *=0{}\ar@{->}[u] &  &  &  & *=0{}\ar@{=}[rrrr] & \ar@{->}[uu] &  & *=0{}\ar@{->}[uu] & *=0{}\\
 &  &  &  &  & *=0{(a)} &  &  &  &  &  &  &  &  & *=0{(b)}
}
\]

\caption{The different diagrams contributing to double parton scattering (DPS)
$(a)$ the two possible $1+2$ processes and $(b)$ a $2+2$ process.
There is no \textquotedblleft$1+1$\textquotedblright{}
contribution as explained in \cite{Blok2012}. the $=$ line represents the hadrons. \label{fig:1+2}}
\end{figure}

Recall that mean field processes  (Fig. \ref{fig:1+2}b) are described by  the mean field theory, based on factorization of $_2GPD$.
The contribution of $1\rightarrow2$  into effective  cross section is usually parametrized  \cite{Blok2014} by a factor
\begin{equation}
R_{DPS}=\frac{\sigma_{eff;1+2}}{\sigma_{eff;2+2}}=2\times\frac{7}{3}\frac{\phantom{}_{\left[1\rightarrow 2\right]}D\left(x_1,x_2,Q_1,Q_2\right)}{G\left(x_1,Q_1\right)G\left(x_2,Q_2\right)}
\end{equation}
so that full effective cross section is 
\begin{equation}
\sigma_{DPS\,\, tot}=\frac{\sigma_{DPS\,\,  mean \,\, field}}{1+R_{DPS}}.
\end{equation}
Here $\phantom{}_{\left[1\rightarrow 2\right]}D$ (denoted as $\phantom{}_{\left[1\right]}D$ in \cite{Blok2014}) is part of the two parton Generalised Parton Distribution $_2GPD$  corresponding to $1\rightarrow 2$ processes
(Fig. \ref{fig:1+2}a), and G is the conventional  PDF.
\par Note that while calculating the contribution of $1\rightarrow  2$ processes we assume that the corresponding 
part of $\phantom{}_2GPD$ denoted by $\phantom{}_{\left[1\rightarrow 2\right]}D$ is calculated at $\vec \Delta=0$, where $\vec \Delta$ is the momentum conjugate to the distance between the 2 partons. This is possible, since the perturbative formfactor,
describing the $\vec \Delta$  dependence of the perturbative $1\rightarrow 2$ vertex decreases 
with $\Delta$  much slower than a nonperturbative two gluon formfactor that describes the $\Delta$ dependence of
$_2$ GPD  \cite{Blok2012}.
In order to compute  $\phantom{}_{\left[1\rightarrow 2\right]}D$ we use the formula given in \cite{Blok2014}

\begin{eqnarray}
\phantom{}_{\left[1\rightarrow 2 \right]}D^{AB}_{h}\left(x_{1},Q_{1},x_{2},Q_{2}\right) & = & \underset{E,A^{\prime},B^{\prime}}{\sum}\intop_{Q_{0}^{2}}^{min\left(Q_{1}^{2},Q_{2}^{2}\right)}\frac{dk^{2}}{k^{2}}\frac{\alpha_{s}\left(k^{2}\right)}{2\pi}\int\frac{dy}{y}G_{h}^{E}\left(y;k^{2}\right)\nonumber \\[10pt]
 & \times & \int\frac{dz}{z\left(1-z\right)}\Phi_{E}^{A^{\prime}}\left(z\right)D_{A^{\prime}}^{A}\left(\frac{x_{1}}{zy};Q_{1}^{2},k^{2}\right)D_{B^{\prime}}^{B}\left(\frac{x_{2}}{\left(1-z\right)y};Q_{2}^{2},k^{2}\right).\nonumber \\[10pt]
\label{eq: PH}
\end{eqnarray}

Here $A$ and $B$ are the parton types and $h$ the original hadron type (from now on we'll only consider gluon distributions inside proton, so we'll suppress these indices) and the sum runs over all processes $E\rightarrow A^{\prime},B^{\prime}$ allowed in leading order.
 $\Phi$ are the DGLAP kernels.
 The functions $D^{B}_{A}\left(x;Q^{2},k^{2}\right)$  are  the fundamental solutions of the DGLAP equation \cite{Dokshitzer1980,Altarelli1977}, that is the probabilities to find a particle $B$ with Bjorken variable $x$ while probing at energy $k^2$ a particle $A$ that has virtuality $Q^2$.
 $Q_0$ is a parameter chosen to separate perturbative and non perturbative contributions to $\phantom{}_2GPD$ and should be between 0.7-1 GeV.

 \par In the same manner one can compute in the leading logarithmic approximation the distribution of three particles coming from two consecutive splittings (Fig. \ref{fig:3+2}c). The leading logarithmic distribution corresponds to the strong ordering of the scales of the consecutive spittings:  $l^2>>k^2$. 
 \par  Unlike the $1\rightarrow2$ this distribution is not fully symmetric because one particle comes from the first splitting and the other two from the second splitting. This distribution has the following form (see  Appendix \ref{sec:Computation-of-the} for details):

\begin{multline}
\phantom{}_{\left[1\rightarrow3\right]}D_{h}^{A;BC}\left(x_{1},Q_{1};x_{2},Q_{2},x_{3},Q_{3}\right)=\\
\underset{E,E^{\prime},E^{\prime\prime},A^{\prime},B^{\prime},C^{\prime}}{\sum}\intop_{Q_{0}^{2}}^{min\left(Q_{1}^{2},Q_{2}^{2},Q_{3}^{2}\right)}\frac{dk^{2}}{k^{2}}\int\frac{dy}{y}G_{h}^{E}\left(y;k^{2}\right)\intop_{k^{2}}^{min\left(Q_{2}^{2},Q_{3}^{2}\right)}\frac{dl^{2}}{l^{2}}\int\frac{dy_{l}}{y_{l}}\times\\
\frac{\alpha_{s}\left(k^{2}\right)}{2\pi}\int\frac{dz}{z\left(1-z\right)}\Phi_{E}^{A^{\prime}}\left(z\right)D_{A^{\prime}}^{A}\left(\frac{x_{1}}{zy};Q_{1}^{2},k^{2}\right)D_{E^{\prime\prime}}^{E^{\prime}}\left(\frac{y_{l}}{\left(1-z\right)y};l^{2},k^{2}\right)\times\\
\frac{\alpha_{s}\left(l^{2}\right)}{2\pi}\int\frac{dz^{\prime}}{z^{\prime}\left(1-z^{\prime}\right)}\Phi_{E^{\prime}}^{B^{\prime}}\left(z^{\prime}\right)D_{B^{\prime}}^{B}\left(\frac{x_{2}}{z^{\prime}y_{l}};Q_{2}^{2},k^{2}\right)D_{C^{\prime}}^{C}\left(\frac{x_{3}}{\left(1-z^{\prime}\right)y_{l}};Q_{3}^{2},k^{2}\right)
\end{multline}

\section{Cross Section of Triple Parton Scattering.}
\subsection{Different Contributions to TPS.  \label{sec:Different}}

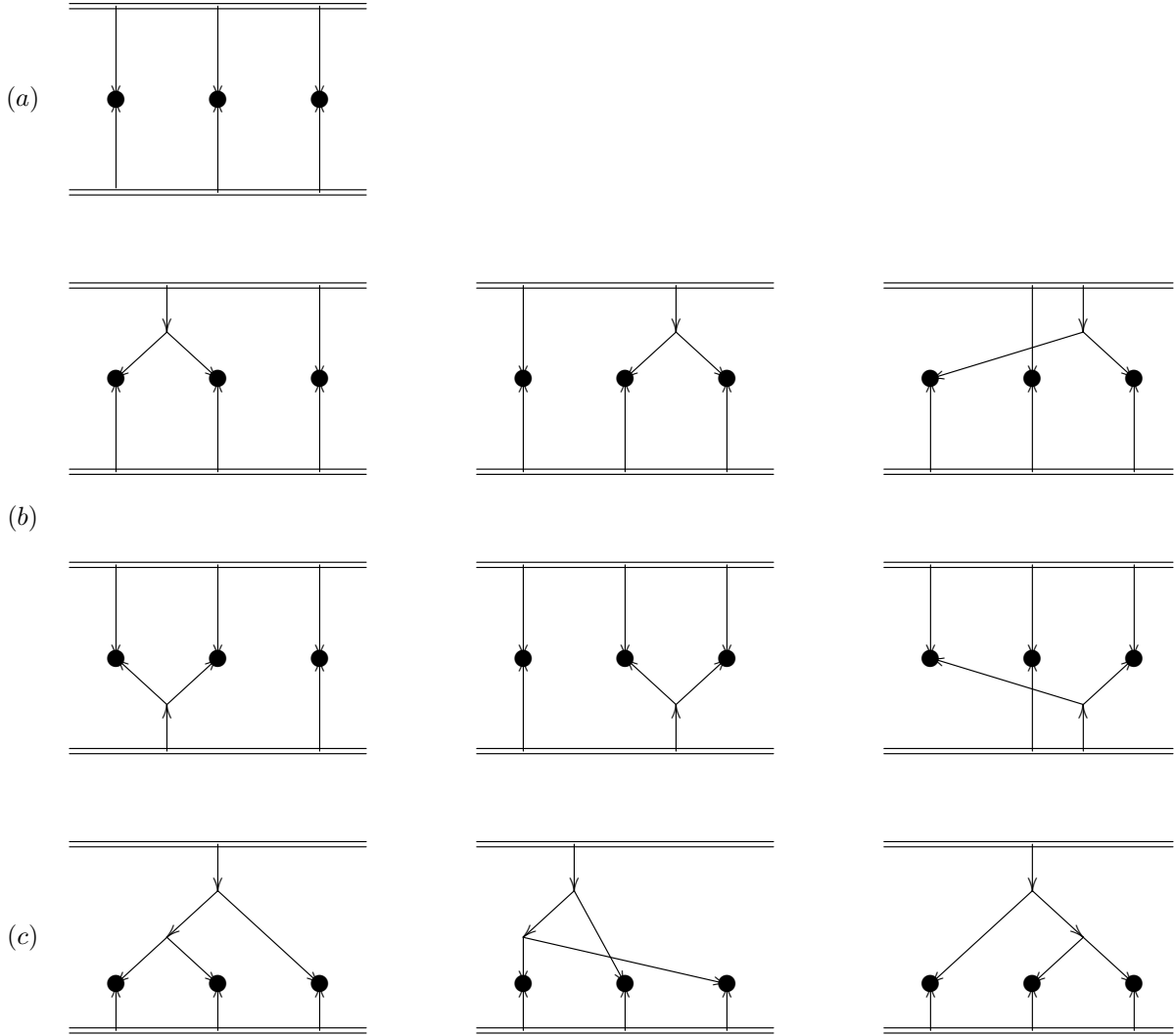
\begin{figure}
\[
\xymatrix{\xyR{1pc}\xyC{1pc} & *=0{}\ar@{=}[rrrrrr] & *=0{}\ar@{->}[dd] &  & *=0{}\ar@{->}[dd] &  & *=0{}\ar@{->}[dd] & *=0{}\\
\\
*=0{(a)} &  & *=0{\newmoon} & {\ } & *=0{\newmoon} & {\ } & *=0{\newmoon}\\
\\
 & *=0{}\ar@{=}[rrrrrr] & \ar@{->}[uu] &  & *=0{}\ar@{->}[uu] &  & *=0{}\ar@{->}[uu] & *=0{}\\
\\
 & *=0{}\ar@{=}[rrrrrr] &  & *=0{}\ar@{->}[d] & *=0{} &  & *=0{}\ar@{->}[dd] & *=0{} &  & *=0{}\ar@{=}[rrrrrr] & *=0{}\ar@{->}[dd] &  &  & *=0{}\ar@{->}[d]*=0{} & *=0{} & *=0{} &  & *=0{}\ar@{=}[rrrrrr] &  &  & *=0{}\ar@{->}[dd] & *=0{}\ar@{->}[d]*=0{} & *=0{} & *=0{}\\
 &  &  & *=0{}\ar@{->}[dl]\ar@{->}[dr] &  &  &  &  &  &  &  &  &  & *=0{}\ar@{->}[dl]\ar@{->}[dr] &  &  &  &  &  &  &  & *=0{}\ar@{->}[dlll]\ar@{->}[dr]\\
 &  & *=0{\newmoon} &  & *=0{\newmoon} & {\ } & *=0{\newmoon} &  & {\ \ } &  & *=0{\newmoon} & {\ } & *=0{\newmoon} &  & *=0{\newmoon} &  &  &  & *=0{\newmoon} & {\ } & *=0{\newmoon} &  & *=0{\newmoon}\\
\\
 & *=0{}\ar@{=}[rrrrrr] & *=0{}\ar@{->}[uu] &  & *=0{}\ar@{->}[uu] &  & *=0{}\ar@{->}[uu] & *=0{} &  & *=0{}\ar@{=}[rrrrrr] & *=0{}\ar@{->}[uu] &  & *=0{}\ar@{->}[uu] & *=0{} & *=0{}\ar@{->}[uu] & *=0{} &  & *=0{}\ar@{=}[rrrrrr] & *=0{}\ar@{->}[uu] &  & *=0{}\ar@{->}[uu] & *=0{} & *=0{}\ar@{->}[uu] & *=0{}\\
*=0{(b)} &  &  &  &  &  &  &  & {\ \ } &  &  &  &  &  &  &  & {\ \ }\\
 & *=0{}\ar@{=}[rrrrrr] & *=0{}\ar@{->}[dd] &  & *=0{}\ar@{->}[dd] &  & *=0{}\ar@{->}[dd] & *=0{} &  & *=0{}\ar@{=}[rrrrrr] & *=0{}\ar@{->}[dd] &  & *=0{}\ar@{->}[dd] &  & *=0{}\ar@{->}[dd] & *=0{} &  & *=0{}\ar@{=}[rrrrrr] & *=0{}\ar@{->}[dd] &  & *=0{}\ar@{->}[dd] &  & *=0{}\ar@{->}[dd] & *=0{}\\
\\
 &  & *=0{\newmoon} & {\ } & *=0{\newmoon} & {\ } & *=0{\newmoon} &  & {\ \ } &  & *=0{\newmoon} & {\ } & *=0{\newmoon} & {\ } & *=0{\newmoon} &  &  &  & *=0{\newmoon} & {\ } & *=0{\newmoon} & {\ } & *=0{\newmoon}\\
 &  &  & *=0{}\ar@{->}[ul]\ar@{->}[ur] &  &  &  &  &  &  &  &  &  & *=0{}\ar@{->}[ul]\ar@{->}[ur] &  &  &  &  &  &  &  & *=0{}\ar@{->}[ulll]\ar@{->}[ur]\\
 & *=0{}\ar@{=}[rrrrrr] &  & *=0{}\ar@{->}[u] &  &  & *=0{}\ar@{->}[uu] & *=0{} &  & *=0{}\ar@{=}[rrrrrr] & *=0{}\ar@{->}[uu] &  &  & *=0{}\ar@{->}[u] &  & *=0{} &  & *=0{}\ar@{=}[rrrrrr] &  &  & *=0{}\ar@{->}[uu] & *=0{}\ar@{->}[u] &  & *=0{}\\
\\
 & *=0{}\ar@{=}[rrrrrr] &  &  & *=0{}\ar@{->}[d] &  & *=0{} & *=0{} &  & *=0{}\ar@{=}[rrrrrr] &  & *=0{}\ar@{->}[d] &  &  & *=0{} & *=0{} &  & *=0{}\ar@{=}[rrrrrr] &  &  & *=0{}\ar@{->}[d] &  & *=0{} & *=0{}\\
 &  &  &  & *=0{}\ar@{->}[dl]\ar@{->}[ddrr] &  &  &  &  &  &  & *=0{}\ar@{->}[dl]\ar@{->}[ddr] &  &  &  &  &  &  &  &  & *=0{}\ar@{->}[dr]\ar@{->}[ddll]\\
*=0{(c)} &  &  & *=0{}\ar@{->}[dl]\ar@{->}[dr] &  &  &  &  & {\ \ } &  & *=0{}\ar@{->}[d]\ar@{->}[drrrr] &  &  &  &  &  & {\ \ } &  &  &  &  & *=0{}\ar@{->}[dr]\ar@{->}[dl]\\
 &  & *=0{\newmoon} &  & *=0{\newmoon} & {\ } & *=0{\newmoon} &  &  &  & *=0{\newmoon} &  & *=0{\newmoon} & {\ } & *=0{\newmoon} &  &  &  & *=0{\newmoon} &  & *=0{\newmoon} & {\ } & *=0{\newmoon}\\
 & *=0{}\ar@{=}[rrrrrr] & *=0{}\ar@{->}[u] &  & *=0{}\ar@{->}[u] &  & *=0{}\ar@{->}[u] & *=0{} &  & *=0{}\ar@{=}[rrrrrr] & *=0{}\ar@{->}[u] &  & *=0{}\ar@{->}[u] &  & *=0{}\ar@{->}[u] & *=0{} &  & *=0{}\ar@{=}[rrrrrr] & *=0{}\ar@{->}[u] &  & *=0{}\ar@{->}[u] &  & *=0{}\ar@{->}[u] & *=0{}\\
\\
}
\]

\caption{The different diagrams contributing to triple parton scattering (TPS): $(a)$ the $3+3$ (or mean field contribution), $(b)$ the $6$ different $2+3$ contributions (other types of $2\rightarrow 3$ diagrams should not contribute in LLA) and $(c)$ the $1+3$ contributions which just like in the $2+3$ have also the "mirror" case (not shown). \label{fig:3+2}}
\end{figure}

The  mean field triple effective cross section defined in (\ref{1}) can be calculated as  \cite{dEnterria:2017yhd}:

\begin{multline}
\frac{1}{\sigma_{eff;TPS}^{2}}=\frac{1}{G\left(x_{1}\right)G\left(x_{2}\right)G\left(x_{3}\right)G\left(x_{1^{\prime}}\right)G\left(x_{2^{\prime}}\right)G\left(x_{3^{\prime}}\right)}\int\frac{d\vec{\Delta}_{1}}{\left(2\pi\right)^{2}}\frac{d\vec{\Delta}_{2}}{\left(2\pi\right)^{2}}\frac{d\vec{\Delta}_{3}}{\left(2\pi\right)^{2}}\left(2\pi\right)^{2}\delta\left(\vec{\Delta}_{1}+\vec{\Delta}_{2}+\vec{\Delta}_{3}\right)\\
\phantom{}_{3}GPD\left(x_{1},\vec{\Delta}_{1},x_{2},\vec{\Delta}_{2},x_{3},\vec{\Delta}_{3}\right)\phantom{}_{3}GPD\left(x_{1^{\prime}},-\vec{\Delta}_{1},x_{2^{\prime}},-\vec{\Delta}_{2},x_{3^{\prime}},-\vec{\Delta}_{3}\right),
\end{multline}

where we suppressed the explicit dependence on the hard scales.
\par  We use the following notations:
\begin{itemize}
\item $x_{1},x_{2},x_{3}$ are the Bjorken variables of the partons from
the first hadron
\item $x_{1^{\prime}},x_{2^{\prime}},x_{3^{\prime}}$ are the Bjorken variables
of the partons from the second hadron
\item $\vec{\Delta}_{i}$ are the conjugate to the distance of the parton
$i$ and $i^{\prime}$ from the center of the hadron
\item $\phantom{}_{3}GPD$ is the generalized 3-parton distribution function 
\end{itemize}
We can write the $\phantom{}_{3}GPD$ as a sum of $3\rightarrow3$ part where all
the partons come from the non-perturbative (NP) wave function of the
hadron, a $2\rightarrow3$ part where 2 partons come from the NP
wavefunction of the hadron but one undergoes a splitting to 2 different
partons and a $1\rightarrow3$  part where 1 parton come from the NP
wavefunction of the hadron and undergoes 2 splittings to form 3 different
partons:

\begin{equation}
\phantom{}_{3}GPD\left(x_{1},\vec{\Delta}_{1},x_{2},\vec{\Delta}_{2},x_{3},\vec{\Delta}_{3}\right)=\phantom{}_{\left[3\rightarrow3\right]}GPD+\phantom{}_{\left[2\rightarrow3\right]}GPD+\phantom{}_{\left[1\rightarrow3\right]}GPD.
\end{equation}

Here we have omitted the explicit dependence of the three parts on $x_i$ and $\vec{\Delta}_{i}$. Using an independent parton approximation for the distribution of
NP partons we write:

\begin{subequations}

\begin{equation}
\phantom{}_{\left[3\rightarrow3\right]}GPD\left(x_{1},\vec{\Delta}_{1},x_{2},\vec{\Delta}_{2},x_{3},\vec{\Delta}_{3}\right)=G\left(x_{1},\vec{\Delta}_{1}\right)G\left(x_{2},\vec{\Delta}_{2}\right)G\left(x_{3},\vec{\Delta}_{3}\right),
\end{equation}

\begin{equation}
\phantom{}_{\left[2\rightarrow3\right]}GPD\left(x_{1},\vec{\Delta}_{1},x_{2},\vec{\Delta}_{2},x_{3},\vec{\Delta}_{3}\right)=\sum_{i\in\left\{ 1,2,3\right\} }G\left(x_{i},\vec{\Delta}_{i}\right)\phantom{}_{\left[1\rightarrow2\right]}D\left(x_{j},\vec{\Delta}_{j},x_{k},\vec{\Delta}_{k}\right).\label{eq:D2->3}
\end{equation}

\begin{equation}
\phantom{}_{\left[1\rightarrow3\right]}GPD\left(x_{1},\vec{\Delta}_{1},x_{2},\vec{\Delta}_{2},x_{3},\vec{\Delta}_{3}\right)=\sum_{i\in\left\{ 1,2,3\right\} }\phantom{}_{\left[1\rightarrow3\right]}D\left(x_{i},\vec{\Delta}_{i};x_{j},\vec{\Delta}_{j},x_{k},\vec{\Delta}_{k}\right).\label{eq:D1->3}
\end{equation}

\end{subequations}

Here  $G\left(x_{1},\vec{\Delta}_{1}\right)$ is the single parton generalized
distribution function (not to be confused with $G\left(x_{1}\right)$  which is the conventional parton distribution function), $\phantom{}_{\left[1\rightarrow2\right]}D$ is  the $1\rightarrow2$ distribution defined in \cite{Blok2014} and  $\phantom{}_{\left[1\rightarrow3\right]}D$ is the $1\rightarrow3$ distribution defined above.
in (\ref{eq:D2->3}) $i$ runs over all possibilities (3) for a parton that does not come from the splitting and $k,l$ are
the other partons. In  Eq. (\ref{eq:D1->3}) the situation is similar, as  $i$ runs over all possibilities (3) for a parton that comes from the first splitting and $k,l$ are
the partons coming from the second splitting.

For simplicity, we now restrict ourselves to the
case $x_{1}=x_{2}=x_{3}=x_{1^{\prime}}=x_{2^{\prime}}=x_{3^{\prime}}$
but the generalisation is straightforward. In fact we are interested only in central kinematics, which 
gives dominant contribution to TPS (see Table 1 in ref. \cite{CMS:2023}).
so this approximation will
be sufficient for numerical calculations in the last section.

In this case, the sum in (\ref{eq:D2->3},\ref{eq:D1->3}) is redundant and we can write without loss
of generality:

\begin{subequations}

\begin{equation}
\phantom{}_{\left[3\rightarrow3\right]}D=G\left(x,\vec{\Delta}_{1}\right)G\left(x,\vec{\Delta}_{2}\right)G\left(x,\vec{\Delta}_{3}\right),
\end{equation}

\begin{equation}
\phantom{}_{\left[2\rightarrow3\right]}D=3\times G\left(x,\vec{\Delta}_{1}\right)\phantom{}_{\left[1\right]}D\left(x,\vec{\Delta}_{2},x,\vec{\Delta}_{3}\right).\label{eq:D2->3-1}
\end{equation}

\begin{equation}
\phantom{}_{\left[1\rightarrow3\right]}D=3\times \phantom{}_{\left[1\rightarrow3\right]}D\left(x,\vec{\Delta}_{1};x,\vec{\Delta}_{2},x,\vec{\Delta}_{3}\right).\label{eq:D1->3-1}
\end{equation}

\end{subequations}

We neglect the $\vec{\Delta}$ dependence of $\phantom{}_{\left[1\rightarrow2\right]}D$ and $\phantom{}_{\left[1\rightarrow3\right]}D$ (as was explained in \cite{Blok2012} for the $1\rightarrow2$ case)
and model that of $G\left(x_{1},\vec{\Delta}_{1}\right)$ by:

\begin{equation}
G\left(x_{1},\vec{\Delta}_{1}\right)=G\left(x_{1}\right)F_{2g}\left(\vec{\Delta}_{1}\right)
\end{equation}

where  $F_{2g}$ is  the 2 gluon form factor, which we model by the  exponential
fit \cite{Blok:2017alw}:

\begin{equation}
F_{2g}\left(\vec{\Delta}_{1}\right)=e^{-\frac{B_{g}}{2}\Delta_{1}^{2}}
\end{equation}
\par Let us comment on the value of parameter $B_g$. The value $B_g\sim4$ $GeV^{-2}$ is extracted from the HERA experiments of exclusive photo production \cite{Frankfurt2002}.  The corresponding mean field effective cross section is  proportional to
$\sim B_g$. We neglect small dependence of $B_g$ on Bjorken $x$ since this dependence effectively cancels out while 
calculating the cross sections.
 Note that for very small transverse momenta the  pQCD effects due to $1\rightarrow 2$ processes
is rather small. So along with \cite{d1} we can take $B_g$ such that the effective mean field cross section at low transverse momenta
is equal to the experimental one, meaning $B_g\sim 0.7-1$ $GeV^{-2}$ leading to  experimentally observed $\sigma_{DPS,\,\,eff}\sim 6-10 $ mb. Our results in this paper will not depend on the particular value of $B_g$ except, of course, the total DPS and TPS cross sections.
\par  We can now write the effective TPS cross section as shown in Fig. \ref{fig:3+2}, where 
we depict all diagrams contributing to DPS in LLA approximation.
\par We use the parametrisation:
\begin{subequations}
\begin{equation}
\frac{1}{\sigma_{eff;TPS}^{2}}=\left[\frac{1}{\sigma_{eff;3+3}^{2}}+\frac{1}{\sigma_{eff;2+3}^{2}}+\frac{1}{\sigma_{eff;1+3}^{2}}\right]=\frac{1}{\sigma_{eff;3+3}^{2}}\left(1+R+R^\prime \right).
\end{equation}
Then
\begin{multline}
\frac{1}{\sigma_{eff;3+3}^{2}}\equiv I_3=\int\frac{d\vec{\Delta}_{1}}{\left(2\pi\right)^{2}}\frac{d\vec{\Delta}_{2}}{\left(2\pi\right)^{2}}\frac{d\vec{\Delta}_{3}}{\left(2\pi\right)^{2}}\left(2\pi\right)^{2}\delta\left(\vec{\Delta}_{1}+\vec{\Delta}_{2}+\vec{\Delta}_{3}\right)\\
F_{2g}\left(\vec{\Delta}_{1}\right)F_{2g}\left(\vec{\Delta}_{2}\right)F_{2g}\left(\vec{\Delta}_{3}\right)F_{2g}\left(-\vec{\Delta}_{1}\right)F_{2g}\left(-\vec{\Delta}_{2}\right)F_{2g}\left(-\vec{\Delta}_{3}\right),
\end{multline}

\begin{multline}
\frac{1}{\sigma_{eff;2+3}^{2}}=2\times3\times\frac{\phantom{}_{\left[1\rightarrow2\right]}D}{G\left(x\right)^{2}}\int\frac{d\vec{\Delta}_{1}}{\left(2\pi\right)^{2}}\frac{d\vec{\Delta}_{2}}{\left(2\pi\right)^{2}}\frac{d\vec{\Delta}_{3}}{\left(2\pi\right)^{2}}\left(2\pi\right)^{2}\delta\left(\vec{\Delta}_{1}+\vec{\Delta}_{2}+\vec{\Delta}_{3}\right)\\
F_{2g}\left(\vec{\Delta}_{1}\right)F_{2g}\left(-\vec{\Delta}_{1}\right)F_{2g}\left(-\vec{\Delta}_{2}\right)F_{2g}\left(-\vec{\Delta}_{3}\right)\equiv2\times3\times\frac{\phantom{}_{\left[1\rightarrow2\right]}D}{G\left(x\right)^{2}}\times I_2,
\end{multline}

\begin{multline}
\frac{1}{\sigma_{eff;1+3}^{2}}=2\times3\times\frac{\phantom{}_{\left[1\rightarrow3\right]}D}{G\left(x\right)^{3}}\int\frac{d\vec{\Delta}_{1}}{\left(2\pi\right)^{2}}\frac{d\vec{\Delta}_{2}}{\left(2\pi\right)^{2}}\frac{d\vec{\Delta}_{3}}{\left(2\pi\right)^{2}}\left(2\pi\right)^{2}\delta\left(\vec{\Delta}_{1}+\vec{\Delta}_{2}+\vec{\Delta}_{3}\right)\\
F_{2g}\left(-\vec{\Delta}_{1}\right)F_{2g}\left(-\vec{\Delta}_{2}\right)F_{2g}\left(-\vec{\Delta}_{3}\right)\equiv2\times3\times\frac{\phantom{}_{\left[1\rightarrow3\right]}D}{G\left(x\right)^{3}}\times I_1,
\end{multline}

\begin{equation}
R=\frac{\sigma_{eff;3+3}^{2}}{\sigma_{eff;2+3}^{2}},\ R^{\prime}=\frac{\sigma_{eff;3+3}^{2}}{\sigma_{eff;1+3}^{2}}.\label{eq:R defenition}
\end{equation}
\end{subequations}

The  factor 2 for $\frac{1}{\sigma_{eff;2/1+3}^{2}}$ comes from the
fact that the splitting can occur in both colliding hadrons, the factors of 3 comes from (\ref{eq:D2->3-1},\ref{eq:D1->3-1}).
For comparison, we'll also compute the equivalent integrals for DPS:

\begin{equation}
I_{2+2}=\int\frac{d\vec{\Delta}_{1}}{\left(2\pi\right)^{2}}\frac{d\vec{\Delta}_{2}}{\left(2\pi\right)^{2}}\left(2\pi\right)^{2}\delta\left(\vec{\Delta}_{1}+\vec{\Delta}_{2}\right)F_{2g}\left(\vec{\Delta}_{1}\right)F_{2g}\left(\vec{\Delta}_{2}\right)F_{2g}\left(-\vec{\Delta}_{1}\right)F_{2g}\left(-\vec{\Delta}_{2}\right),
\end{equation}

\begin{equation}
I_{1+2}=\int\frac{d\vec{\Delta}_{1}}{\left(2\pi\right)^{2}}\frac{d\vec{\Delta}_{2}}{\left(2\pi\right)^{2}}\left(2\pi\right)^{2}\delta\left(\vec{\Delta}_{1}+\vec{\Delta}_{2}\right)F_{2g}\left(-\vec{\Delta}_{1}\right)F_{2g}\left(-\vec{\Delta}_{2}\right).
\end{equation}

\subsection{Calculation of TPS  contributions.\label{sec:calculations}}
We shall now calculate different contributions to TPS scattering and see that their ratios are essentially determined  by geometric factors.
 \par The integrals $I_3,I_2,I_1,I_{2+2},I_{1+2}$  are Gaussian and can be easily calculated using the standard formula:
 \begin{equation}
 \int d^2x_1..d^2x_n Exp(-\sum A_{ij}x_ix_j)=\pi^n/det(A)
 \end{equation}
 where A is a rectangular matrix. We immediately obtain:

\begin{equation}
I_{3}
=  \frac{1}{\left(2\pi\right)^{2}}\frac{1}{12B_{g}^{2}}\,\,\,,
I_{2} 
=  \frac{1}{\left(2\pi\right)^{2}}\frac{1}{5B_{g}^{2}}\,\,\,,
I_{1} 
=  \frac{1}{\left(2\pi\right)^{2}}\frac{1}{3B_{g}^{2}}.
\end{equation}

\par For the DPS integrals we have in the same way:

\begin{equation}
I_{2+2}=  \frac{1}{8\pi B_{g}}\, , \,I_{1+2}=\frac{1}{4\pi B_{g}}.
\end{equation}

In order to check the mean field calculations we define:

\begin{equation}
k=\frac{\sigma_{eff;3+3}^{TPS}}{\sigma_{eff;2+2}^{DPS}}=\frac{I_{3}^{-\frac{1}{2}}}{I_{2+2}^{-1}}=\frac{2\pi\sqrt{12}B_{g}}{8\pi B_{g}}=\frac{\sqrt{12}}{4}\approx0.86\ ,
\end{equation}
which is consistent with similar ratios defined in \cite{dEnterria:2017yhd}. We now can write $R$ as defined in (\ref{eq:R defenition}). For simplicity we shall not write indices $1,2,3$ explicitly, so i.e.
\begin{equation}
G^2\equiv G(x_1,Q_1)\cdot G(x_2,Q_2).
\end{equation}
\par Although we do not write the arguments explicitly, $G$ and $\phantom{}_{\left[1\rightarrow2/3\right]}D$ depend on the kinematics of the process (i.e the Bjorken variable $x$ and the hard scale $Q$) as well as the types of the participating partons. 

\begin{equation}
R=2\times3\times\frac{I_{2}}{I_{3}}\frac{\phantom{}_{\left[1\rightarrow2\right]}D}{G^{2}}=6\times\frac{12}{5}\frac{\phantom{}_{\left[1\rightarrow2\right]}D}{G^{2}}\approx14.4\cdot\frac{\phantom{}_{\left[1\right]}D}{G^{2}},\label{eq:R}
\end{equation}

this can be compared to the DPS case for which 
\begin{equation}
R_{DPS}=2\times\frac{7}{3}\frac{\phantom{}_{\left[1\right]}D}{G^{2}}\approx4.6\cdot\frac{\phantom{}_{\left[1\right]}D}{G^{2}},
\end{equation}

which means that the enhancement of the cross section from parton splitting
for TPS is more than 3 times that of DPS. On the other hand, the ratio $R^\prime$ will get a geometrical factor of:

\begin{equation}
R^\prime=2\times3\times\frac{I_{1}}{I_{3}}\times\frac{\phantom{}_{\left[1\rightarrow3\right]}D}{G\left(x\right)^{3}}=2\times3\times\frac{12}{3}\frac{\phantom{}_{\left[1\rightarrow3\right]}D}{G\left(x\right)^{3}}=24\cdot\frac{\phantom{}_{\left[1\rightarrow3\right]}D}{G\left(x\right)^{3}}.\label{eq:Rprime}
\end{equation}

We'll explain how to compute $\phantom{}_{\left[1\rightarrow3\right]}D$ in more detail in Appendix \ref{sec:Computation-of-the}.

\section{Numerical  Results. \label{sec:numeric}}

\begin{figure}

\includegraphics[scale=1]{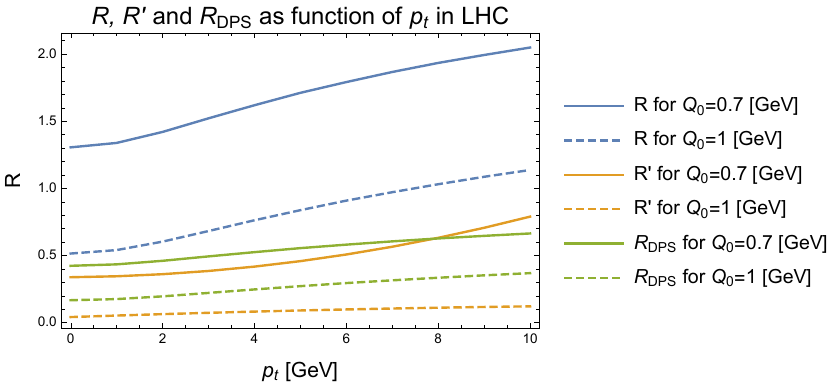}
\caption{The ratios $R=\frac{\sigma_{eff;3+3}^{2}}{\sigma_{eff;2+3}^{2}}$, $R^{\prime}=\frac{\sigma_{eff;3+3}^{2}}{\sigma_{eff;1+3}^{2}}$ and $R_{DPS}=2\times\frac{7}{3}\frac{\phantom{}_{\left[1\rightarrow2\right]}D}{G^{2}}$. We remember that $G$ and $\phantom{}_{\left[1\right]}D$ depend on $p_t$ through their dependence on both $Q$ and $x$ as is explained in the text.
 \label{R}}
\end{figure}

\par Our results imply that the ratio of TPS to DPS effective cross sections is given by
\begin{equation}
\frac{\sigma_{eff;TPS}}{\sigma_{eff;DPS}}=k \cdot \frac{\sqrt{1+R+R^{\prime}}}{1+ R_{DPS}}\label{43}.
\end{equation}

\begin{figure}

\includegraphics[scale=1]{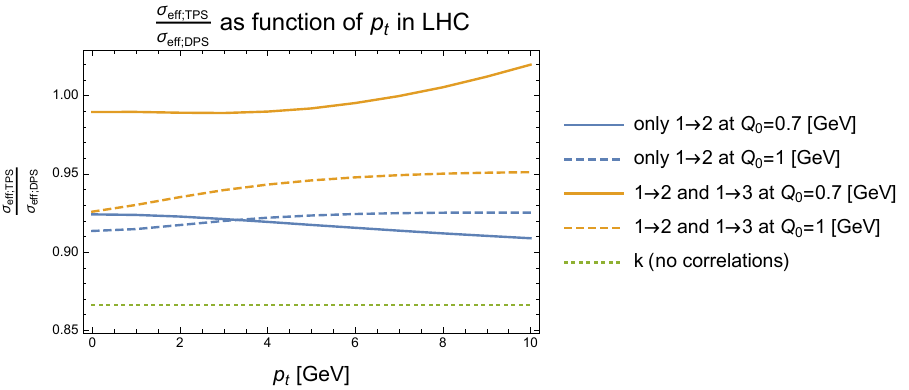}\caption{The ratio $\frac{\sigma_{eff;TPS}}{\sigma_{eff;DPS}}$ \label{D to T ratio}}

\end{figure}

\par  The inclusion of  $1\rightarrow2$ processes makes  the ratio $\sigma_{TPS}/\sigma_{`DPS}$,  as well as each of them separately,
dependent on transverse scale. The characteristic virtuality of charmonium is \cite{kopfj,Vogt:2017osu}
\begin{equation}
Q^2\sim p_t^2+4m^2_c
\end{equation}
where $p_t$ is its transverse momentum and $m_c=1.5\  [GeV]$.
In actual situations, the transverse momentum is $\ge 3.5-6$ GeV and less than 10 GeV \cite{CMS:2023}.
Then, we have in our kinematics
\begin{equation}
x_1=x_2=x=\sqrt{\frac{4p_t^2}{s}}\sim\sqrt{\frac{4Q^2}{s}}.
\end{equation}
Here  $s$ is the center of mass energy of the hadron collision. In our calculations, we'll work in the LHC kinematics for which  $s=1.96\times10^{8}\ \left[GeV^{2}\right]$. (see \cite{Ellis1996} for the details of kinematics in 2 to 2 hard collisions).
\par Since the dominant contribution comes from the central kinematics (i.e small rapidity) \cite{CMS:2023} we can neglect  the rapidity dependence and assume rapidity 
$\eta =0$ \cite{Blok2014,Blok:2017alw}.
The transverse dependence of $R$, $R^\prime$ and in comparison $R_{DPS}$ is depicted in Fig. \ref{R}.
 Our results for the ratio $\frac{\sigma_{eff;TPS}}{\sigma_{eff;DPS}}$ as a function of transverse momenta are depicted in Fig. \ref{D to T ratio}.
\begin{figure}

\includegraphics[scale=1]{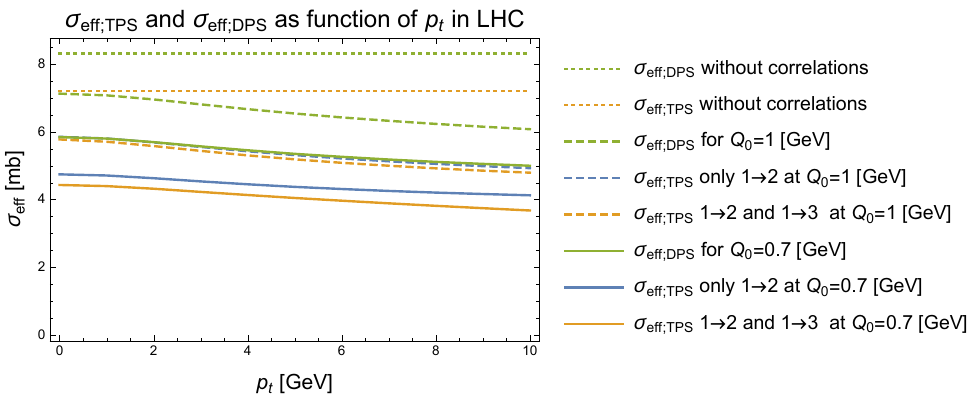}\caption{$\sigma_{eff;TPS}$ and $\sigma_{eff;DPS}$ as a function of $p_t$ with and without the contribution of correlations \label{Sigmas}}

\end{figure}
We see that  the ratio  $\frac{\sigma_{eff;TPS}}{\sigma_{eff;DPS}}$  depends on transverse momenta only moderately, but when taking all processes into account can change from the no correlation value of $k\sim 0.85$  to values of order 1, i.e.   up to 15 perecent.
 
\begin{figure}

\includegraphics[scale=1]{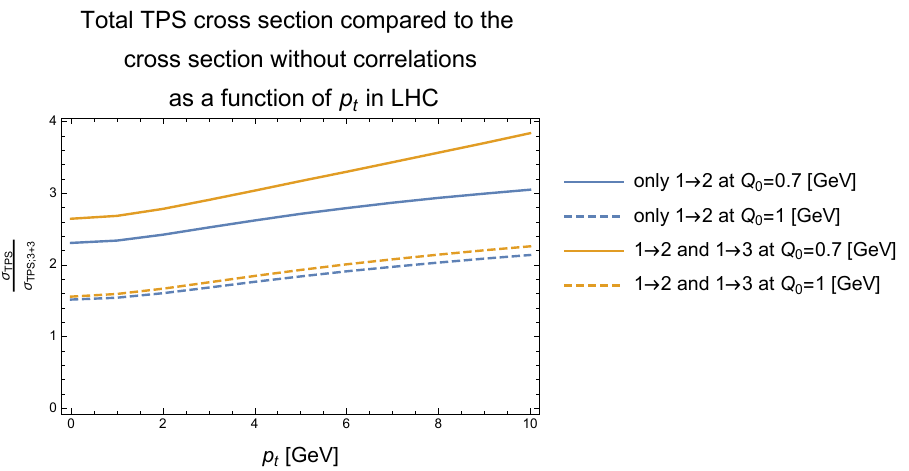}\caption{The ratio of the TPS cross section computed with correlations to the TPS cross section computed only in mean field \label{TPS Sigma}}

\end{figure}

On the other hand Figs. \ref{Sigmas}, \ref{TPS Sigma} show  that the $1\rightarrow2$  and  $1\rightarrow3$ processes lead to a dependence of the total effective cross sections on charmonium transverse scale $p_t$.

\par Our results in Figs. \ref{R}, \ref{D to T ratio} may be  important for actual experimental determination of the TPS cross section. The actual separation
between DPS and TPS  may be influenced by  inclusion of the $p_t$ dependence of the TPS and DPS effective cross sections.
\par  Our results depicted in Figs. \ref{Sigmas}, \ref{TPS Sigma} describe the dependence of the corresponding effective cross sections on $p_t$.
It will be very interesting to check the current and future experimental data if such dependence indeed exists,
in distinction from the mean field approach, where these effective cross sections are model independent.
\acknowledgements The authors are indebtful to M. Strikman for reading the article.
The research was supported by an ISF grant 2025311 and BSF grant 2020115.

\appendix

\section{Computation of the $1\rightarrow3$ Process\label{sec:Computation-of-the} }

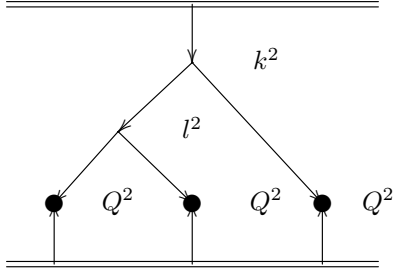
\begin{figure}
\[
\xymatrix{*=0{}\ar@{=}[rrrrrr] &  &  & *=0{}\ar@{->}[d] &  & *=0{} & *=0{}\\
 &  &  & *=0{}\ar@{->}[dl]\ar@{->}[ddrr] & {k^{2}}\\
 &  & *=0{}\ar@{->}[dl]\ar@{->}[dr] & {l^{2}}\\
 & *=0{\newmoon} & {Q^{2}} & *=0{\newmoon} & {Q^{2}} & *=0{\newmoon} & {Q^{2}}\\
*=0{}\ar@{=}[rrrrrr] & *=0{}\ar@{->}[u] &  & *=0{}\ar@{->}[u] &  & *=0{}\ar@{->}[u] & *=0{}
}
\]

\caption{An illustration of the $1\rightarrow3$ process with the scale for
each splittings (and the scale evolution) shown exactly: The first
splitting at  the scale $k$ the second at the scale $l$, the hard process
scale is $Q$. \label{1to3}}
\end{figure}

The computation of the $1\rightarrow3$ distribution is very similar
to the one of the $1\rightarrow2$ distribution. In fact, it's just
taking a $1\rightarrow2$ distribution as the ``initial condition''
of another $1\rightarrow2$ distribution, those creating 2 splittings
to form the $1\rightarrow3$ distribution. For this to be true the second splitting scale must be much larger than the first, in order to create the needed  logarithm in the evolution.

 The $1\rightarrow2$ distribution
is given in (\ref{eq: PH}), so replacing the hadron PDF $G$ with
another $1\rightarrow2$ distribution we have:

\begin{align}
\phantom{}_{\left[1\rightarrow3\right]}D_{h}^{ABC}\left(x_{1},Q_{1};x_{2},Q_{2},x_{3},Q_{3}\right) & =\nonumber \\
 & \underset{E,B^{\prime},C^{\prime}}{\sum}\intop_{Q_{0}^{2}}^{min\left(Q_{2}^{2},Q_{3}^{2}\right)}\frac{dk^{2}}{k^{2}}\frac{\alpha_{s}\left(k^{2}\right)}{2\pi}\int\frac{dy}{y}\phantom{}_{\left[1\rightarrow2\right]}D_{h}^{AE}\left(x_{1},Q_{1},y,k\right)\times\nonumber \\
 & \int\frac{dz}{z\left(1-z\right)}\Phi_{E}^{B^{\prime}}\left(z\right)D_{B^{\prime}}^{B}\left(\frac{x_{2}}{zy};Q_{2}^{2},k^{2}\right)D_{C^{\prime}}^{C}\left(\frac{x_{3}}{\left(1-z\right)y};Q_{3}^{2},k^{2}\right).
\end{align}

Or more explicitly changing the names of the second splitting from
$k\rightarrow l$, $y\rightarrow y_{l}$, $E\rightarrow E^{\prime}$
and $z\rightarrow z^{\prime}$ we have:

\begin{multline}
\phantom{}_{\left[1\rightarrow3\right]}D_{h}^{ABC}\left(x_{1},Q_{1};x_{2},Q_{2},x_{3},Q_{3}\right)=\\
\underset{E,E^{\prime},E^{\prime\prime},A^{\prime},B^{\prime},C^{\prime}}{\sum}\intop_{Q_{0}^{2}}^{min\left(Q_{1}^{2},Q_{2}^{2},Q_{3}^{2}\right)}\frac{dk^{2}}{k^{2}}\int\frac{dy}{y}G_{h}^{E}\left(y;k^{2}\right)\intop_{k^{2}}^{min\left(Q_{2}^{2},Q_{3}^{2}\right)}\frac{dl^{2}}{l^{2}}\int\frac{dy_{l}}{y_{l}}\times\\
\frac{\alpha_{s}\left(k^{2}\right)}{2\pi}\int\frac{dz}{z\left(1-z\right)}\Phi_{E}^{A^{\prime}}\left(z\right)D_{A^{\prime}}^{A}\left(\frac{x_{1}}{zy};Q_{1}^{2},k^{2}\right)D_{E^{\prime\prime}}^{E^{\prime}}\left(\frac{y_{l}}{\left(1-z\right)y};l^{2},k^{2}\right)\times\\
\frac{\alpha_{s}\left(l^{2}\right)}{2\pi}\int\frac{dz^{\prime}}{z^{\prime}\left(1-z^{\prime}\right)}\Phi_{E^{\prime}}^{B^{\prime}}\left(z^{\prime}\right)D_{B^{\prime}}^{B}\left(\frac{x_{2}}{z^{\prime}y_{l}};Q_{2}^{2},k^{2}\right)D_{C^{\prime}}^{C}\left(\frac{x_{3}}{\left(1-z^{\prime}\right)y_{l}};Q_{3}^{2},k^{2}\right).
\end{multline}

This process together, with the splitting scales, is shown explicitly in Fig. \ref{1to3}. These integrals can be numerically computed just like for the $\phantom{}_{\left[1\rightarrow2\right]}D_{h}^{AE}$
case. It should be noted that the $1\rightarrow2$ distribution receives
its biggest contribution for small $Q_{0}<k\ll Q$. The $1\rightarrow3$
distribution have very limited phase space for both $k$ and $l$
and therefore should give significant contributions only if $Q_{0}$
is to be small enough to allow for 2 splitting in the region of
$Q_{0}<k\ll l \ll Q$ . This is indeed verified in Fig. \ref{TPS Sigma}
where it can be seen that the $1\rightarrow3$ contribution is much
smaller at $Q_{0}=1\ [GeV]$ but is rather large for $Q_{0}=0.7\ [GeV]$.

\end{document}